\documentclass[proceedings]{JHEP} 

\conference{Quantum aspects of gauge theories, supersymmetry and unification,
Paris 1999}

\title{Wilson loop via AdS/CFT duality\footnote{Talk given by S.\ F\"orste.}}

\author{Stefan F\"orste$^a$, Debashis Ghoshal$^b$ and Stefan Theisen$^c$\\
        $^a$ Institute of Physics, Bonn University, Nu\ss allee 12,
        Bonn 53115, Germany\\
       $^b$ Metha Research Institute of Mathematics \& Mathematical
        Physics, Chhatnag Road, Jhusi, Allahabed 211019, India\\
       $^c$ Sektion Physik, Universit\"at M\"unchen, 
        Theresienstra\ss e 37, 80333 M\"unchen, Germany\\   
        E-mail: \email{forste@th.physik.uni-bonn.de,
        ghoshal@mri.ernet.in,
        theisen@theorie.physik.uni-muenchen.de}}

\abstract{
The Wilson loop in ${\cal N}=4$ supersymmetric Yang-Mills theory admits 
a dual description as a macroscopic string configuration in the adS/CFT 
correspondence. We discuss the correction to the quark anti-quark 
potential arising from the fluctuations of the superstring. }

\begin{document}

\section{Introduction}
One important ingredient of the string dualities is the twofold 
description of D-branes as solitonic supergravity solutions
and as submanifolds of spacetime where open strings may
end. The second  description leads to a gauge field 
theory on the world-volume of D-branes. Based on this general idea is 
Maldacena's conjecture\cite{maldacena} relating superconformal field
theories to  supergravity (or superstrings) living in a higher
dimensional space with boundary. (For a review containing also a
comprehensive list of references, see \cite{aharony}.) 
For example, ${\cal N }=4$ 
supersymmetric $SU(N)$ Yang Mills theory in four dimensions is dual 
to the type IIB string theory on $adS_5\times S^5$ space. The radii 
of the $S^5$ and the $adS_5$ are equal and related to the Yang-Mills
coupling via
$R^2/\alpha^\prime = \sqrt{4\pi g^2_{YM}N}$. (The string coupling $g$
is $g=g^2_{YM}$.) For supergravity to be a good effective description, 
we need the radius of curvature to be large and also the string 
coupling to be small. This means that we need $g_s$ small but 
$g_s N$ large. 
The latter is however the 't Hooft coupling constant in 
the large $N$ field theory which is thus strongly coupled. 
Maldacena's conjecture provides a new possibility to gain
insight into strongly coupled gauge theory by studying 
weakly coupled string theory. 
As an application, Wilson loops have been computed in 
\cite{rey-yee} and \cite{mal-wil}. 
The string configuration for a quark-antiquark pair
separated by a distance $L$, 
is a long string on $adS_5\times S^5$, the ends of which are
restricted to the (four dimensional) boundary of $adS_5$, where
they are at a distance $L$ apart. The expectation value of the Wilson 
loop is then given by the effective energy of the string. We will
review this computation in the next section. In the third section
leading corrections are discussed. As a further application of our
techniques we briefly review in section four 
membrany corrections to the Wilson
surface in M5 theory. We conclude with a short summary.

\section{Review}

The dual description of ${\cal N}=4$ super Yang-Mills theory is a
type IIB string living in $adS_5\times S^5$. In particular the target
space metric ($G_{MN}$) is 
\begin{eqnarray}
ds^2 & = & R^2\left[ U^2\left( - \left(dx^0\right)^2 + \left(dx^1\right)^2
    + \left(dx^2\right)^2 \right.\right. \nonumber\\
& & \left. \left. +\left(dx^3\right)^2\right)+ \frac{dU^2}{U^2}
    + d\Omega_5 ^2\right].\label{metric}
\end{eqnarray}
Compared to \cite{mal-wil} we have rescaled $U\rightarrow R^2U$,
and also set $\alpha'=1$. In addition there is a constant dilaton and $N$
units of RR-4-form flux through $S^5$. In order to compute the Wilson
loop one minimizes the Nambu-Goto 
action \cite{mal-wil,rey-yee}
\footnote{After switching off the world-sheet fermions, 
the type IIB action reduces to the Nambu-Goto action.}
\begin{equation}\label{classical}
S_{NG} = \int \frac{d^2\sigma }{2\pi}\sqrt{-\det G_{MN}\partial_i X^M
  \partial_j X^N}
\end{equation}
with the boundary condition that the ends of the string are separated
by a distance $L$ at $U=\infty$.
We work in the static gauge ($X^0 =\tau$, $X^1=\sigma$) and restrict to
the case that all coordinates but $U$ are constant. The radial
coordinate $U$ depends on $\sigma$. Then the (implicit)
solution reads\footnote{In the following we will just
  take the upper sign with the understanding that the square root
  stands for both branches.}
\begin{equation}\label{background}
\partial_\sigma U = \pm \frac{U^2}{U_0 ^2}\sqrt{U^4 - U_0 ^4} ,
\end{equation}
where 
\begin{equation} \label{solution}
U_0 =
\frac{\left(2\pi\right)^\frac{3}{2}}{\Gamma\left(\frac{1}{4}\right)^2 L
  } 
\end{equation}
is the minimal $U$-value the string obtains.
The energy of the quark-antiquark pair is the length
of the geodesic (open string) connecting them. One finds (after subtracting 
an $L$ independent infinite contribution
from the self energy of the heavy quarks)
\begin{equation}\label{wilsonvev}
E= - \frac{4\pi\sqrt{g_{YM}^2N}}{\Gamma\left(1/4\right)^4L} .
\end{equation}
This strong coupling result differs from the perturbative field theory 
computation ($g_{YM}^2N$ small), which predicts a Coulomb law with 
$E\sim g^2_{YM}N/L$. In general the numerator is a function of 
$g^2_{YM}N$ which interpolates between these two forms, and hence
there ought to be corrections to (\ref{wilsonvev}) which is the 
result of a classical supergravity computation. 
Since $adS_5\times S^5$ is an exact string background
\cite{banks,MeTs,kallosh}, the first correction comes from the 
fluctuations of the superstring ($R^2/\alpha'$ correction). In this talk 
we will report on work in this direction\cite{us}. Corrections due to 
string fluctuations have been discussed in \cite{oleson}, and
subsequently in \cite{Naik, Zarembo, Sonne, Kinar, Gross}.
Ref.\cite{Erickson} considered corrections to the field theoretical 
result.

\section{Fluctuations}

The quantum theory of type IIB 
string in this background is described by the action in \cite{MeTs}.
It is a Green-Schwarz type sigma model action on a target supercoset.
The usual sigma model expansion in $R^2/\alpha'$ results in a power 
series in $1/\sqrt{Ng_{YM}^2}$. Since conformal invariance prevents 
the appearance of a new scale these corrections are not expected to 
change the $1/L$ dependence of $E$ on dimensional grounds. However
they can modify the `Coulomb charge'.

The leading order correction is 
obtained by expanding around the classical configuration (\ref{background})
to second order in fluctuations. We parameterize the bosonic fluctuations 
by normal coordinates\cite{afm}: $\xi^a$ on $adS_5\times S^5$, (here 
$a=0,\cdots,4;5,\cdots,9$ are local (flat) Lorentz indices; 
$\xi^4$ is in the $U$ direction). 
Using the 
normal coordinate expansion one ensures that the functional measure for 
the fluctuations is translation invariant. This takes care of potential
problems due to the curved target space. Ref. \cite{Gross}
has an extensive discussion on additional subtleties in the functional 
measures due to a curved world sheet. 

At second order, the bosonic fluctuations in $adS_5$ and 
$S^5$ directions and the fermionic fluctuations decouple. 
Before writing their action, we define the combinations
\begin{eqnarray}
\xi^\parallel & = & \frac{U_0^2}{U^2}\xi^1 +\frac{\sqrt{U^4 -
    U_0^4}}{U^2}\xi^4,\nonumber\\
\xi^\perp &=& -\frac{\sqrt{U^4-U_0^4}}{U^2}\xi^1 +
    \frac{U_0^2}{U^2}\xi^4,\label{parallel}
\end{eqnarray}
which parameterize fluctuations along the longitudinal
and perpendicular directions of the background string in the one-four plane.
Now the $adS_5$ part of the action becomes
\begin{eqnarray} 
\!\! S_{adS}^{(2)} & = &\frac{1}{4\pi}\int d^2\sigma\sqrt{-h}
\Bigg[ h^{ij}\left(\sum_{a=2,3,\perp}\! \partial_i \xi^a\partial_j \xi^a 
\right)\nonumber\\
\! &+ &\! 2\left(\xi^2\right)^2 + 2\left(\xi^3\right)^2\! +
  2\frac{U^4\!\! -U_0^4}{U^4}\left(\xi^\perp\right)^2\Bigg] \label{ads-fluc} 
\end{eqnarray}
where $h_{ij}$ is (up to a factor of $R^2$) the classical induced
world-sheet metric\footnote{For our purpose it is more convenient to
work with the induced metric rather than the standard (conformally)
flat one on the world-sheet.}
\begin{equation}\label{2dmetric}
ds^2 = -U^2 \left( d\sigma^0\right)^2 +
\frac{U^6}{U_0^4}\left(d\sigma^1\right)^2 .
\end{equation}
Observe that $\xi^0$ and $\xi^\parallel$ do not appear in (\ref{ads-fluc})
(total derivatives have been dropped). Hence a natural choice to fix 
world-sheet diffeomorphisms is
\begin{equation} \label{lc-gauge}
\xi^0 = \xi^\parallel = 0 .
\end{equation}
The action quadratic in fluctuations in $S^5$ directions comes out to
be
\begin{equation} \label{s5-fluc}
S_{S^5}^{(2)} = \frac{1}{4\pi} \int d^2\sigma \sqrt{-h} h^{ij}
\sum_{a=5}^9 \partial_i\xi^a \partial_j \xi^a.
\end{equation}

The fermionic part of the action is given by plugging in the background 
(\ref{solution}) into the action of \cite{MeTs} and keeping terms quadratic 
in fermions. This action has local fermionic $\kappa$-symmetry which has
to be fixed for the one-loop calculation. There is a proposal in the
literature\cite{KaTs} to this end. For our purpose, however, it turns
out that the following choice is most convenient.\footnote{We will
  comment on a different choice below.} We will set (in the
notation of \cite{MeTs})
\begin{equation}\label{kappa-fix}
\left(\gamma^-\right)^\alpha_{\,\beta}
\theta^{1\beta\beta^\prime}=0\quad,
\quad\left(\gamma^+\right)^\alpha _{\,\beta}
\theta^{2\beta\beta^\prime}=0\quad 
\end{equation}
where $\gamma^{\pm} = \gamma^0 \pm \gamma^\parallel$ with
$\gamma^\parallel = \frac{U_0^2}{U^2}\gamma^1 +
\frac{\sqrt{U^4-U_0^4}}{U^2}\gamma^4$ (Cf.\ ({\ref{parallel})). 
With this choice the target space spinors `metamorphose' into world-sheet
spinors, and the action is found to simplify substantially. The
corresponding equations of motion are most compactly written for the
`two-component' world-sheet spinors $\left(\begin{array}{c}\theta^1\\
\theta^2\end{array}\right)$:
\begin{equation}\label{fermioneom}
\left[ i\left( \rho^m\nabla_m\right)^\alpha_{\,\beta} -
  \delta^\alpha_\beta\rho^3\right]\left(
\begin{array}{c} \theta^{1\beta\alpha'} \\
  \theta^{2\beta\alpha'}\end{array}\right) =0 .
\end{equation}
The notation needs explanation. Firstly, the covariant derivatives
act as
\begin{eqnarray}
\left(\nabla_\pm\theta^I\right)^{\alpha\alpha^\prime}& = & \nonumber
\\ & &
\!\!\!\!\!\!\!\!\!\!\!\!\!\!\!\!\!\!\!\!\!
\left[ \delta^\alpha _\beta\left(\partial_\pm \pm
  \frac{\omega_\pm}{2}\right)+\left(A_\pm\right)^\alpha_{\,\beta}\right]
  \theta^{I\beta\alpha^\prime}, 
\end{eqnarray}
where the tangent space derivatives
\begin{equation} \label{tderivs}
\partial_\pm={1\over U}\partial_\tau\pm{U_0^2\over U^3}\partial_\sigma
\end{equation}
are defined with the help of a (inverse) zweibein $\epsilon_m$ of 
the metric (\ref{2dmetric}), $\omega_m^{01}=\epsilon^\tau_m\omega_\tau^{01}$
being the corresponding spin connection. There is an additional gauge
field 
\begin{equation} \label{gaugefield}
A_\pm = \pm\frac{U_0^2}{U^2}\gamma^{14}
\end{equation}
for local rotations in 
the tangent one-four-plane. Finally, the matrices
\begin{equation}
\rho^+ = \left( \begin{array}{cc} 0 & 0\\ \! \gamma^0 & 0
  \end{array}\right)\quad ,\quad \rho^- = \left(
\begin{array}{cc}0 & \! \gamma^0\\ 0 & 0\end{array}\right)
\end{equation}
satisfy a two dimensional Clifford algebra, and 
$\rho^3=[\rho^+,\rho^-]$. The fermionic action is easily inferred from
the equations of motion (\ref{fermioneom}).

Collecting our results (\ref{ads-fluc}), (\ref{s5-fluc}) and
(\ref{fermioneom}) one can write a formal expression for the 1-loop
contribution to the effective action as a ratio of determinants
of two dimensional generalized Laplace operators\cite{us}. These
determinants suffer from divergences and can be
regularized by, say, the heat kernel technique\cite{gilkey}. 
The quadratic divergences cancel, but naively a logarithmic 
divergence remains, which may be absorbed in the (infinite) mass 
of the external quarks\cite{us}. As unsatisfactory as it may be, it
does not affect our result for the correction to the Coulomb charge.
More recently the authors of \cite{Gross} have argued that this
divergence should, as in flat space, actually cancel.
As far as the corrections to the Coulomb charge are concerned the
results of \cite{Gross} and ours \cite{us} are actually equivalent. 
In the 
following we demonstrate that the apparently different expressions
for the fermionic operators in \cite{us} and \cite{Gross} are related
to each other by a local Lorentz rotation\footnote{From the sigma
  model point of view this is just a field redefinition with unit Jacobian.}.
To this end, define
\begin{equation} \label{redef}
\theta^{I} = S \psi^{I} ,
\end{equation}
where we suppressed target space spinor indices. For the matrix $S$ we 
choose the one given in Ref.\cite{Gross}, Section 6.3,
\begin{equation}
S= \cos \frac{\alpha}{2} -\sin \frac{\alpha}{2}\gamma^{14},
\end{equation}
where
\begin{eqnarray}
\cos\alpha & = & \frac{U_0^2}{U^2},\\
\sin \alpha & = & \frac{\sqrt{U^4-U_0^4}}{U^2},
\end{eqnarray}
implying that
\begin{equation}
\partial_\sigma \alpha = 2U .
\end{equation}
The Dirac operator for $\psi$ is given by conjugation by $S$ from
the one for $\theta$. Using the fact that $S$ commutes with the $\rho_m$ 
and using ((\ref{tderivs}), (\ref{gaugefield}))
\begin{equation}
S^{-1}A_\pm S = - S^{-1}\partial_\pm S = \pm \frac{U_0 ^2}{U^2}\gamma^{14},
\end{equation}
we find that the Dirac operator acting on the fields $\psi^I$
(\ref{redef}) takes again the form (\ref{fermioneom}), but now the
connection $A_\pm$ has been gauged away. It is also easy to show that
for the redefined spinors $\psi ^I$
the kappa-fixing condition takes the form (\ref{kappa-fix}), but with
\begin{equation}
\gamma^\parallel \rightarrow S^{-1}\gamma^\parallel S =\gamma^1 ,
\end{equation}
which is the same as (6.35) in \cite{Gross}. This shows that the
results in \cite{Gross} are equivalent to ours.

\section{Wilson surface in M5 theory}

The Maldacena conjecture also applies to the case of M-theory living on
$adS_7 \times S^4$ being dual to the field theory on a stack of M5
branes\cite{maldacena}. The metric of $adS_7 \times S^4$ is
\begin{equation}
ds^2 = l_p ^2 R^2\left[ U^2 \eta_{\mu\nu}dx^\mu dx ^\nu +
  4\frac{dU^2}{U^2} + d\Omega_4^2\right],
\end{equation}
where we have rescaled the five-brane coordinates $x^\mu$ by $R^{3/2}$
as compared to\cite{maldacena}, $d\Omega_4 ^2$ is the metric on $S^4$.
In the limit that the eleven dimensional Planck length $l_p$ goes to zero
M-theory on $adS_7\times S^4$ is conjectured to be dual to the field
theory on $N$ M5-branes, where the adS radius and the number of
five-branes are related,
\begin{equation}
R=\left( \pi N\right)^{\frac{1}{3}} .
\end{equation}
Higher curvature corrections will be small as long as $N$ is taken to
be large. In analogy to the previously discussed Wilson loop one can
study a Wilson surface in M5-theory. The set up is a membrane
extending along the $x^{0,1,2}$ and the $U$ direction ending in two
parallel lines separated by a distance $L$ at the boundary of
$adS_7$\cite{mal-wil}. In the following we will recall this set-up
(with slightly changed conventions) and thereafter study corrections
due to fluctuations of the membrane. This will be a brief summary of
the work presented in\cite{me}.
The classical background corresponding to the Wilson surface is
obtained by minimizing the world volume of the membrane
\begin{equation}
S= \frac{1}{2\pi}\int \sqrt{-\det \left( G_{MN}\partial_a
    X^M\partial_b X^N\right)}
\end{equation}
with appropriate boundary conditions. The indices $M,N$ label the
eleven target space coordinates and $a,b$ are world volume coordinates
($\tau, \sigma , \phi$).
By choosing the static gauge $X^0 = \tau$, $x^1 = \sigma$, $X^2 =
\phi$ and assuming all other coordinates but $U= U\left(\sigma\right)$
to be constant one finds the solution,
\begin{equation}
\partial_\sigma U = \pm \frac{U^2}{2U_0^3}\sqrt{U^6 -U_0 ^6} .
\end{equation}
Requiring that the membrane ends in two parallel strings at distance
$L$ determines the integration constant
\begin{equation}
U_0 = \frac{2}{3L}B\left(\frac{2}{3}, \frac{1}{2}\right),
\end{equation}
where $B$ denotes Euler's beta-function. Computing the vacuum energy density
of this configuration one obtains (again after subtracting an $L$
independent infinite contribution to the self-energy of the strings)
the potential between the two strings living on the M5-branes,
\begin{equation}
\varepsilon_{pot} = - \frac{2R^3}{27\pi} B\left
  ( \frac{2}{3},\frac{1}{2}\right)^3 \frac{1}{L^2} .
\end{equation}
This result is reliable for large $N$ where the geometry is not
corrected and the classical approximation dominates. In
\cite{Kallosh:1998qs} it was argued that there are no corrections to
the geometry due to finite $N$. Another potential source for
corrections are fluctuations of the membrane around its classical
background described above. Again we expand in normal
coordinates\cite{afm} and obtain the action quadratic in bosonic
fluctuations. We trade the fluctuations in one- and
six-direction\footnote{The fluctuations are labeled by Lorentz
  indices; $\xi^6$ is in the $U$ direction.} for normal and parallel ones,
\begin{eqnarray}
\xi^\parallel & = & \frac{U_0^3}{U^3}\xi^1 + \frac{\sqrt{U^6 - U_0
    ^6}}{U^3} \xi^6 \\
\xi^\perp & = & -\frac{\sqrt{U^6 - U_0
    ^6}}{U^3} \xi^1 +\frac{U_0^3}{U^3}\xi^6 .
\end{eqnarray}
The contribution from the $adS$ directions is
\begin{eqnarray}
S^{(2)}_{adS} & = & \frac{1}{4\pi}\int d^3 \sigma\sqrt{-h}\left
  [ h^{ij}\sum_{a=3}^{5,\perp}\partial_i\xi^a\partial_j\xi^a\right.
  \nonumber\\
& &\!\!\!\!\!\!\!\!\!\!\!\! 
+ \left.\frac{3}{4} \sum_{a=3}^{5}\left(\xi^a\right)^2
  +\left(\frac{9}{4} - R^{(3)}\right)\left(\xi^\perp\right)^2\right]
\label{ads7-fluc}
\end{eqnarray}
where $h_{ij}$ is (up to a factor of $R^2$) the induced metric,
\begin{equation} \label{induced-m}
ds^2 = - U^2 d\tau^2 + \frac{U^8}{U_0^6}d\sigma^2 + U^2 d\phi^2  ,
\end{equation}
and $R^{(3)}$ is the corresponding scalar curvature,
\begin{equation}
R^{(3)}=\frac{3}{2}\frac{U^6 + U_0^6}{U^6} .
\end{equation}
Again the three longitudinal directions $0,2,\parallel$ drop out of
the action and we gauge them to zero. The bosonic fluctuations in
$S^4$ direction have a simple action,
\begin{equation} \label{s4-fluc}
S_S^{(2)} = \frac{1}{4\pi}\int d^3\sigma
\sqrt{-h}h^{ij}\sum_{a=7}^{10}\partial_i\xi^a\partial_j\xi^a .
\end{equation}
In order to discuss the fermionic fluctuations we take the $\kappa$
symmetric action of\cite{deWit:1998yu}. For our background the part
quadratic in fermions consists out of terms containing $\Gamma^a$
($a=0,\ldots 
,6$), where $\Gamma^a$ is an eleven dimensional gamma matrix. These
can be written as $\Gamma^a = \gamma^a \otimes \gamma^{5^\prime}$ where
the lower case gammas are gamma matrices in the tangent spaces of
$adS_7$ and $S^4$, respectively. We split the 32-component spinors
into two 16-component spinors ($\theta^1$, $\theta^2$) according to
their eigenvalue with 
respect to $\gamma^{5^\prime}$. ($\gamma^{5^\prime}\theta^I = -
(-)^I\theta^I$.) A convenient kappa-fixing condition turns out be
\begin{equation} \label{m-kappa-fix}  
\left(1 - (-1)^I\gamma^{0\parallel 2}\right)\theta^I = 0 ,
\end{equation}
where now $\gamma^\parallel = \frac{U_0 ^3}{U^3}\gamma^1 + \frac{\sqrt{U^6
  -U_0^6}}{U^3}\gamma^6 $ . Imposing the kappa-fixing condition
we find that the equations of motion for e.g.\ $\theta^1$ are
\begin{equation}\label{fermions}
\rho^ae_a ^i\left( \partial_i + \frac{1}{4}\omega_i^{bc}\rho_{bc} 
+ A_i\right)\theta^1 = - \frac{3}{4}\theta^1,
\end{equation}
where $e^i _a$ and $\omega_i^{bc}$ are the vielbeine and
spin-connections computed from (\ref{induced-m}) (for details
see \cite{me}). The matrices $\rho$ are
\begin{equation} 
\rho^0 =\gamma^0,\, \rho^1 =\gamma^{02}, \, \rho^2 = \gamma^2
\end{equation}
satisfying a 3d Clifford algebra. The field $A_\sigma =
\frac{3U}{4}\gamma^{16}$ is a background value for a gauge connection
belonging to local rotations in the 1-6 plane. (For $\theta^2$ one
obtains the same result but with a minus sign in the definition of $\rho^1$.) 
Note that the Dirac operator appearing in (\ref{fermions}) is covariant
from a world volume perspective.
Collecting the results (\ref{ads7-fluc}), (\ref{s4-fluc}) and
(\ref{fermions}) one can express the contribution to the energy
density in terms of determinants of covariant operators. These can be
analyzed using for example the heat-kernel method\cite{gilkey}. In
difference to the previously discussed string case one finds divergent
contributions as well to the self-energy density as to the potential
energy density\cite{me}. It would be interesting to investigate 
whether one can
extend the arguments of \cite{Gross} to cancel those
divergences. (Since the discussion in \cite{Gross} is quite heavily
based on conformal invariance and 2d calculus this is not
straightforward.) 
Finally, let us point out that also in the membrane case one can gauge
away the connection appearing in (\ref{fermions}). This can be
achieved by a field redefinition $\theta^I = S\psi^I$ with 
\begin{equation}
S= \cos \frac{\alpha}{2} - \sin\frac{\alpha}{2}\gamma^{16},
\end{equation}
where
\begin{eqnarray}
\cos \alpha & = & \frac{U_0 ^3}{U^3} ,\nonumber \\
\sin \alpha & = & \frac{\sqrt{U^6 - U_0 ^6}}{U^3} .
\end{eqnarray}
The kappa-fixing condition is again (\ref{m-kappa-fix}) but with 
$\gamma^\parallel$ replaced by $\gamma^1$. This coincides with the
kappa-fixing condition advertised in\cite{Claus:1999fh}.

\section{Summary}
In this talk we presented techniques for computing stringy corrections
to the Wilson loop in ${\cal N} =4$ supersymmetric Yang-Mills theory.
The final result can be expressed in terms of determinants of 2d
covariant operators. A result for the leading correction to the
Coulomb charge in terms of a number is still missing (a rough
estimate can be found in \cite{Gross}). We commented on
the relation between our results and those obtained in \cite{Gross}.
In the end we reviewed the application of our techniques to the case
of a Wilson surface in M5-theory. There the result is less
satisfactory as divergences also affect the Coulomb charge.
\vskip0.5cm
\noindent
{\bf Acknowledgments}

\smallskip
\noindent
This work was supported in part by GIF (German Israeli
Foundation for Scientific Research) and the EC programs
ERB-FMRX-CT-96-0045 and 96-0090. D.G.\ acknowledges the support of the 
Humboldt Foundation. 
S.F. and S.T.
thank the organizers of the Paris meeting for creating a pleasant 
and stimulating atmosphere during the conference.


\end{document}